\documentclass[11pt]{article}
\usepackage{amsmath,amssymb,latexsym}

\advance \topmargin by -\headheight
\advance \topmargin by -\headsep
     
\evensidemargin \oddsidemargin
\newcommand\rf[1]{(\ref{eq:#1})}
\newcommand\lab[1]{\label{eq:#1}}
\newcommand\nonu{\nonumber}
\newcommand\br{\begin{eqnarray}}
\newcommand\er{\end{eqnarray}}
\newcommand\be{\begin{equation}}
\newcommand\ee{\end{equation}}

\newcommand\foot[1]{\footnotemark\footnotetext{#1}}
\newcommand\lb{\lbrack}
\newcommand\rb{\rbrack}

\newcommand\llb{\left\lbrack}
\newcommand\rrb{\right\rbrack}

\renewcommand\({\left(}
\renewcommand\){\right)}
\renewcommand\v{\vert}                     
\newcommand\bgv{\bigg\vert}              

\newcommand\bc{\begin{center}}
\newcommand\ec{\end{center}}

\newcommand\Tr{\mathop{\mathrm Tr}}                  
\newcommand\partder[2]{\frac{{\partial {#1}}}{{\partial {#2}}}}

\renewcommand\d{\delta}

\newcommand\vareps{\varepsilon}
\newcommand\g{\gamma}
\newcommand\G{\Gamma}

\newcommand\h{\frac{1}{2}}
\renewcommand\k{\kappa}
\renewcommand\l{\lambda}
\renewcommand\L{\Lambda}
\newcommand\m{\mu}
\newcommand\n{\nu}
\renewcommand\o{\over}

\newcommand\p{\phi}
\newcommand\vp{\varphi}
\renewcommand\P{\Phi}
\newcommand\pa{\partial}

\newcommand\pr{\prime}

\renewcommand\r{\rho}
\newcommand\s{\sigma}
\renewcommand\S{\Sigma}
\renewcommand\t{\tau}
\renewcommand\th{\theta}

\newcommand\wti{\widetilde}
\newcommand\twomat[4]{\left(\begin{array}{cc}  
{#1} & {#2} \\ {#3} & {#4} \end{array} \right)}

\newcommand\cA{{\mathcal A}}

\newcommand\cF{{\mathcal F}}

\newcommand\cL{{\mathcal L}}
\newcommand\cM{{\mathcal M}}

\newcommand{\ct}[1]{\cite{#1}}
\newcommand{\bib}[1]{\bibitem{#1}}
\newcommand\NPB[3]{(#2), \textsl{Nucl. Phys.} \textbf{B#1} #3}

\newcommand\PRD[3]{(#2), \textsl{Phys. Rev.} \textbf{D#1} #3}

\newcommand\PLB[3]{(#2), \textsl{Phys. Lett.} \textbf{#1B} #3}
\newcommand\CQG[3]{(#2), \textsl{Class. Quantum Grav.} \textbf{#1} #3}

\newcommand\IJMPA[3]{(#2), \textsl{Int. J. Mod. Phys.} \textbf{A#1} #3}

\begin{document}

\title{Weyl-Conformally Invariant $p$-Brane Theories}

\author{Eduardo Guendelman and Alexander Kaganovich\\
\small\it Department of Physics, Ben-Gurion University, Beer-Sheva, Israel  \\[-1.mm]
\small\it email: guendel@bgumail.bgu.ac.il, alexk@bgumail.bgu.ac.il \\
${}$ \\
Emil Nissimov and Svetlana Pacheva\\
\small\it Institute for Nuclear Research and Nuclear Energy,\\[-1.mm]
\small\it Bulgarian Academy of Sciences, Sofia, Bulgaria  \\[-1.mm]
\small\it email: nissimov@inrne.bas.bg, svetlana@inrne.bas.bg}
\date{ }
\maketitle

\begin{abstract}
We discuss in some detail the properties of a novel class of Weyl-conformally
invariant $p$-brane theories which describe intrinsically light-like branes for any
odd world-volume dimension and whose dynamics significantly differs from that
of the ordinary (conformally non-invariant) Nambu-Goto $p$-branes. We present 
explicit solutions of the \textsl{WILL}-brane (Weyl-Invariant Light-Like brane)
equations of motion in various gravitational backgrounds of physical
relevance exhibiting the following new phenomena: (i) In spherically symmetric
static backgrounds the \textsl{WILL}-brane automatically positions itself on
(materializes) the event horizon of the corresponding black hole solutions, 
thus providing an explicit dynamical realization of the membrane paradigm in 
black hole physics; (ii) In product spaces (of interest in Kaluza-Klein
context) the \textsl{WILL}-brane wrappes non-trivially around the compact
(internal) dimensions and moves as a whole with the speed of light in the 
non-compact (space-time) dimensions.
\end{abstract}

\maketitle

\section{Introduction}

Higher-dimensional extended objects ($p$-branes, $Dp$-branes) play an
increasingly crucial role in modern non-perturbative string theory of
fundamental interactions at ultra-high energies (for a background on string and 
brane theories, see refs.\ct{brane-string-rev}). Their importance stems primarily 
from such basic properties as: providing explicit realization of non-perturbative
string dualities, microscopic description of black-hole physics, gauge 
theory/gravity correspondence, large-radius compactifications of extra dimensions, 
cosmological brane-world scenarios in high-energy particle phenomenology, 
\textit{etc.} .

In an independent development new classes of field theory models involving gravity,
based on the idea of replacing the standard Riemannian integration measure
(Riemannian volume-form) with an alternative non-Riemannian volume-form or,
more generally, employing on equal footing both Riemannian and non-Riemannian 
volume-forms, have been proposed few years ago \ct{TMT-basic}. Since then,
these new models called \textsl{two-measure theories} have been a subject of 
active research and applications \ct{TMT-recent} \foot{For related ideas, see 
\ct{Hehl-etal}.}.
Two-measure theories address various basic problems in cosmology
and particle physics, and provide plausible solutions for a broad array of issues, 
such as:
scale invariance and its dynamical breakdown; spontaneous generation of
dimensionfull fundamental scales;  
the cosmological constant problem;  
the problem of fermionic families; 
applications to dark energy problem and modern cosmological brane-world scenarios.
For a detailed discussion we refer to the series of papers \ct{TMT-basic,TMT-recent}.

Subsequently, the idea of employing an alternative non-Riemannian integration
measure was applied systematically to string, $p$-brane and $Dp$-brane models
\ct{m-string}. The main feature of these new classes of modified
string/brane theories is the appearance of the pertinent string/brane
tension as an additional dynamical degree of freedom beyond the usual string/brane
physical degrees of freedom, instead of being introduced \textsl{ad hoc} as
a dimensionfull scale. The dynamical string/brane tension acquires the
physical meaning of a world-sheet electric field strength (in the string
case) or world-volume $(p+1)$-form field strength (in the $p$-brane case) and
obeys Maxwell (Yang-Mills) equations of motion or their higher-rank
antisymmetric tensor gauge field analogues, respectively. As a result of the
latter property the modified-measure string model with dynamical tension
yields a simple classical mechanism of ``color'' charge confinement. 

One of the drawbacks of modified-measure $p$-brane and $Dp$-brane models,
similarly to the ordinary Nambu-Goto $p$-branes, is that Weyl-conformal
invariance is lost beyond the simplest string case ($p\! =\! 1$). On the
other hand, it turns out that the form of the action of the modified-measure string
model with dynamical tension suggests a natural way to construct explicitly a 
radically new class of {\em Weyl-conformally invariant} $p$-brane models 
{\em for any} $p$ \ct{will-brane-kiten}. The most profound property of the 
latter models is that for any even $p$ they describe the dynamics of inherently 
{\em light-like} $p$-branes which makes them significantly different
both from the standard Nambu-Goto (or Dirac-Born-Infeld) branes as well as from 
their modified versions with dynamical string/brane tensions \ct{m-string} 
mentioned above.

Before proceeding to the main exposition, which is the detailed discussion
of the properties of the new Weyl-conformally invariant light-like branes,
let us briefly recall the standard Polyakov-type formulation of the ordinary 
(bosonic) Nambu-Goto $p$-brane action:
\be
S = -{T\o 2}\int d^{p+1}\s\,\sqrt{-\g}\, \Bigl\lb 
\g^{ab}\pa_a X^\m \pa_b X^\n G_{\m\n}(X) - \L (p-1)\Bigr\rb \; .
\lab{stand-brane-action}
\ee
Here $\g_{ab}$ is the ordinary Riemannian metric on the $p+1$-dimensional
brane world-volume with $\g \equiv \det\v\v \g_{ab}\v\v$. The world-volume indices
$a,b=0,1,\ldots ,p$ ; 
~$G_{\m\n}$ denotes the Riemannian metric in the embedding space-time with 
space-time indices $\m,\n=0,1,\ldots ,D-1$.
$T$ is the given \textsl{ad hoc} brane tension; the constant
$\L$ can be absorbed by rescaling $T$ (see below Eq.\rf{stand-brane-action-NG}).
The equations of motion w.r.t. $\g^{ab}$ and $X^\m$ read:
\be
T_{ab} \equiv \( \pa_a X^\m \pa_b X^\n - 
\h \g_{ab} \g^{cd}\pa_c X^\m \pa_d X^\n \) G_{\m\n} + \g_{ab} {\L\o 2}(p-1)
= 0 \; ,
\lab{stand-brane-gamma-eqs}
\ee
\be
\pa_a \(\sqrt{-\g}\g^{ab}\pa_b X^\m\) +
\sqrt{-\g}\g^{ab}\pa_a X^\n \pa_b X^\l \G^\m_{\n\l} = 0  \; ,
\lab{stand-brane-X-eqs}
\ee
where:
\be
\G^\m_{\n\l}=\h G^{\m\k}\(\pa_\n G_{\k\l}+\pa_\l G_{\k\n}-\pa_\k G_{\n\l}\)
\lab{affine-conn}
\ee
is the Cristoffel connection for the external metric.

Eqs.\rf{stand-brane-gamma-eqs} when $p \neq 1$ imply:
\be
\L \g_{ab} = \pa_a X^\m \pa_b X^\n G_{\m\n} \; ,
\lab{stand-brane-metric-eqs}
\ee
which in turn allows to rewrite Eq.\rf{stand-brane-gamma-eqs} as:
\be
T_{ab} \equiv \( \pa_a X^\m \pa_b X^\n - 
{1\o {p+1}} \g_{ab} \g^{cd}\pa_c X^\m \pa_d X^\n \) G_{\m\n} = 0 \; .
\lab{stand-brane-gamma-eqs-Pol}
\ee
Furthermore, using \rf{stand-brane-metric-eqs} the Polyakov-type brane action 
\rf{stand-brane-action} becomes on-shell equivalent to the Nambu-Goto-type brane
action:
\be
S = - T \L^{-{{p-1}\o 2}} \int d^{p+1}\s\,
\sqrt{-\det\v\v \pa_a X^\m \pa_b X^\n G_{\m\n} \v\v} \; .
\lab{stand-brane-action-NG}
\ee

Let us note the following properties of standard Nambu-Goto $p$-branes
manifesting their crucial differences w.r.t. the Weyl-conformally invariant
branes discussed below.  Eq.\rf{stand-brane-metric-eqs} tells us that: (i) the 
induced metric on the Nambu-Goto $p$-brane world-volume is {\em non-singular};
(ii) standard Nambu-Goto $p$-branes describe intrinsically {\em massive} modes.

\section{String and Brane Models with a Modified World-Sheet/World-Volume 
Integration Measure}

Here we briefly recall the construction of modified string and ($p$- and
$Dp$-)brane models with dynamical tension based on the use of alternative
non-Riemannian world-sheet/world-volume volume form (integration measure
density) \ct{m-string}.

The modified-measure bosonic string model is given by the following action:
\be
S = - \int d^2\s\,\P (\vp) \Bigl\lb \h\g^{ab} \pa_a X^{\m} \pa_b X^{\n}G_{\m\n}(X)
- \frac{\vareps^{ab}}{2\sqrt{-\g}} F_{ab}(A)\Bigr\rb 
+ \int d^2\s\,\sqrt{-\g} A_a J^a 
\lab{m-string}
\ee
with the notations:
\be
\P (\vp) \equiv \h \vareps_{ij} \vareps^{ab} \pa_a \vp^i \pa_b \vp^j
\quad ,\quad F_{ab} (A) = \pa_{a} A_{b} - \pa_{b} A_{a} \; ,
\lab{m-string-notat}
\ee
$\g_{ab}$ denotes the intrinsic Riemannian world-sheet metric with 
$\g = \det\Vert\g_{ab}\Vert$~ and $G_{\m\n}(X)$ is the Riemannian metric of
the embedding space-time ($a,b=0,1; i,j=1,2; \m,\n =0,1,\ldots,D-1$).

Below is the list of differences w.r.t. the standard Nambu-Goto string
(in the Polyakov-like formulation) :
\begin{itemize}
\item
New non-Riemannian integration measure density $\P (\vp)$ built in terms of
auxiliary world-sheet scalar fields $\vp^i$ ($i=1,2$), independent of the
world-sheet metric $\g_{ab}$, instead of the standard Riemannian one $\sqrt{-\g}$;
\item
Dynamical string tension $T \equiv \frac{\P (\vp)}{\sqrt{-\g}}$ instead of
\textsl{ad hoc} dimensionfull constant;
\item
Auxiliary world-sheet gauge field $A_a$ in a would-be ``topological'' term
$\int d^2\s\, \frac{\P (\vp)}{\sqrt{-\g}} \h\vareps^{ab} F_{ab}(A)$;
\item
Optional natural coupling of auxiliary $A_a$ to external conserved world-sheet 
electric current $J^a$ (see last term in \rf{m-string} and 
Eq.\rf{Maxwell-like-eqs} below).
\end{itemize}

The modified string model \rf{m-string} is Weyl-conformally invariant
similarly to the ordinary case. Here Weyl-conformal symmetry is given by 
Weyl rescaling of $\g_{ab}$ supplemented with a special diffeomorphism in 
$\vp$-target space:
\be
\g_{ab} \longrightarrow \g^{\pr}_{ab} = \rho\,\g_{ab}  \quad ,\quad
\vp^{i} \longrightarrow \vp^{\pr\, i} = \vp^{\pr\, i} (\vp) 
\;\; \mathrm{with} \;\; 
\det \Bigl\Vert \frac{\pa\vp^{\pr\, i}}{\pa\vp^j} \Bigr\Vert = \rho \; .
\lab{Weyl-conf}
\ee

The dynamical string tension appears as a canonically conjugated momentum
w.r.t. $A_1$: 
$\pi_{A_1} \equiv \partder{\cL}{\dot{A_1}} = \frac{\P (\vp)}{\sqrt{-\g}}
\equiv T$, \textsl{i.e.}, $T$ has the meaning of a 
\textit{world-sheet electric field strength}, and the equations of motion w.r.t. 
auxiliary gauge field $A_a$ look exactly as $D=2$ Maxwell eqs.:
\be
\frac{\vareps^{ab}}{\sqrt{-\g}}\pa_b T + J^a = 0 \; .
\lab{Maxwell-like-eqs}
\ee
In particular, for $J^a= 0$ :
\be
\vareps^{ab} \pa_{b} \Bigl(\frac{\P (\vp)}{\sqrt{-\g}}\Bigr) = 0 \qquad,\quad
\frac {\P (\vp)}{\sqrt{-\gamma}}  \equiv T = \textrm{const}\; ,
\lab{Maxwell-like-eqs-0}
\ee
one gets a {\em spontaneously induced} constant string tension.
Furthermore, when the modified string couples to point-like charges on the 
world-sheet (\textsl{i.e.}, $J^0 {\sqrt{-\gamma}} = \sum_i e_i \d (\s - \s_i)$
in \rf{Maxwell-like-eqs}) one obtains classical charge {\em confinement}: 
$\sum_i e_i = 0$.

The above charge confinement mechanism has also been generalized in \ct{m-string}
to the case of coupling the modified string model with dynamical tension to 
non-Abelian world-sheet ``color'' charges. The latter is achieved as follows.
Notice the following identity in $2D$ involving Abelian gauge field $A_a$:
\be
\frac{\vareps^{ab}}{2\sqrt{-\g}} F_{ab}(A) = 
\sqrt{-\h F_{ab}(A) F_{cd}(A) \g^{ac}\g^{bd}} \; .
\lab{2D-id}
\ee 
Then the extension of the action \rf{m-string} to the non-Abelian case is
straightforward:
\be
S = - \int d^2\s \,\P (\vp) \Bigl\lb \h \g^{ab} \pa_a X^{\m} \pa_b X^\n
G_{\m\n}(X) - \sqrt{-\h \Tr (F_{ab}(A)F_{cd}(A)) \g^{ac}\g^{bd}}\Bigr\rb
+ \int d^2\s\,\Tr \( A_a j^a\)
\lab{m-string-NA}
\ee
with $F_{ab}(A) = \pa_a A_b - \pa_b A_c + i \bigl\lb A_a,\, A_b\bigr\rb$,
sharing the same principal property -- dynamical generation of string
tension as an additional degree of freedom.

Similar construction has also been proposed for higher-dimensional
modified-measure $p$- and $Dp$-brane models whose brane tension appears as
an additional dynamical degree of freedom. On the other hand, like the 
standard Nambu-Goto branes, they are Weyl-conformally {\em non}-invariant and 
describe dynamics of {\em massive} modes.

\section{Weyl-Invariant Branes: Action and Equations of Motion}

The identity \rf{2D-id} suggests how to construct \textbf{Weyl-invariant} 
$p$-brane models for any $p$. Namely, we consider the following novel class of
$p$-brane actions:
\be
S = - \int d^{p+1}\s \,\P (\vp) 
\Bigl\lb \h \g^{ab} \pa_a X^{\m} \pa_b X^{\n} G_{\m\n}(X)
- \sqrt{F_{ab}(A) F_{cd}(A) \g^{ac}\g^{bd}}\Bigr\rb
\lab{WI-brane}
\ee
\be
\P (\vp) \equiv \frac{1}{(p+1)!} \vareps_{i_1\ldots i_{p+1}} 
\vareps^{a_1\ldots a_{p+1}} \pa_{a_1} \vp^{i_1}\ldots \pa_{a_{p+1}} \vp^{i_{p+1}}
\; ,
\lab{mod-measure-p}
\ee
where notations similar to those in \rf{m-string} are used 
(here $a,b=0,1,\ldots,p; i,j=1,\ldots,p+1$).

The above action \rf{WI-brane} is invariant under Weyl-conformal symmetry 
(the same as in the dynamical-tension string model \rf{m-string}):
\be
\g_{ab} \longrightarrow \g^{\pr}_{ab} = \rho\,\g_{ab}  \quad ,\quad
\vp^{i} \longrightarrow \vp^{\pr\, i} = \vp^{\pr\, i} (\vp) 
\;\; \mathrm{with} \;\; 
\det \Bigl\Vert \frac{\pa\vp^{\pr\, i}}{\pa\vp^j} \Bigr\Vert = \rho \; .
\lab{Weyl-conf-p}
\ee
  
Let us note the following significant differences of \rf{WI-brane} w.r.t. the standard
Nambu-Goto $p$-branes (in the Polyakov-like formulation) :
\begin{itemize}
\item
New non-Riemannian integration measure density $\P (\vp)$ instead of $\sqrt{-\g}$,
and {\em no}~ ``cosmological-constant'' term ($(p-1)\sqrt{-\g}$);
\item
Variable brane tension $\chi \equiv \frac{\P (\vp)}{\sqrt{-\g}}$ 
which is Weyl-conformal {\em gauge dependent}: $ \chi \to \rho^{\h(1-p)}\chi$;
\item
Auxiliary world-volume gauge field $A_a$ in a ``square-root'' Maxwell
(Yang-Mills) term\foot{``Square-root'' Maxwell (Yang-Mills) action in $D=4$
was originally introduced in the first ref.\ct{Spallucci} and later
generalized to ``square-root'' actions of higher-rank
antisymmetric tensor gauge fields in $D\geq 4$ in the second and third
refs.\ct{Spallucci}.};
the latter is straightforwardly generalized to the non-Abelian case 
-- $\sqrt{-\Tr \( F_{ab}(A) F_{cd}(A)\) \g^{ac}\g^{bd}}$ similarly to
\rf{m-string-NA};
\item
Natural optional couplings of the auxiliary gauge field $A_a$ to external 
world-volume ``color'' charge currents $j^a$;
\item
The action \rf{WI-brane} is manifestly Weyl-conformal invariant for {\em any} $p$;
it describes {\em intrinsically light-like} $p$-branes for any even $p$, as
it will be shown below.
\end{itemize}

In what follows we shall frequently use the short-hand notations: 
\be
\(\pa_a X \pa_b X\) \equiv \pa_a X^\m \pa_b X^\n G_{\m\n}\quad ,\quad
\sqrt{FF\g\g} \equiv \sqrt{F_{ab} F_{cd} \g^{ac}\g^{bd}} \; .
\lab{short-hand}
\ee
Employing \rf{short-hand} the equations of motion w.r.t. measure-building 
auxiliary scalars $\vp^i$ and w.r.t. $\g^{ab}$ read, respectively:
\be
\h \g^{cd}\(\pa_c X \pa_d X\) - \sqrt{FF\g\g} = M \; \Bigl( = \mathrm{const}\Bigr)
\; ,
\lab{phi-eqs}
\ee
\be
\h\(\pa_a X \pa_b X\) + \frac{F_{ac}\g^{cd} F_{db}}{\sqrt{FF\g\g}} = 0 \; ,
\lab{gamma-eqs}
\ee
Taking the trace in \rf{gamma-eqs} implies $M=0$ in Eq.\rf{phi-eqs}.

Next we have the following equations of motion w.r.t. auxiliary gauge field $A_a$
and w.r.t. $X^\m$, respectively:
\be
\pa_b \(\frac{F_{cd}\g^{ac}\g^{bd}}{\sqrt{FF\g\g}} \P (\vp)\) = 0 \; ,
\lab{A-eqs}
\ee
\be
\pa_a \(\P (\vp) \g^{ab}\pa_b X^\m\) +
\P (\vp) \g^{ab}\pa_a X^\n \pa_b X^\l \G^\m_{\n\l} = 0 \; ,
\lab{X-eqs}
\ee
where $\G^\m_{\n\l}$ is the Cristoffel  connection corresponding to the external 
space-time metric $G_{\m\n}$ as in \rf{affine-conn}.

The $A_a$-equations of motion \rf{A-eqs} can be solved in terms of $(p-2)$-form gauge 
potentials $\L_{a_1\ldots a_{p-2}}$ dual w.r.t. $A_a$. The respective
field-strengths are related as follows:
\br
F_{ab}(A)= -\frac{1}{\chi}\,\frac{\sqrt{-\g}\,\vareps_{abc_1\ldots c_{p-1}}}{2(p-1)}
\g^{c_1 d_1}\ldots \g^{c_{p-1} d_{p-1}} 
\, F_{d_1\ldots d_{p-1}}(\L) \,\g^{cd} \(\pa_c X \pa_d X\) \; ,
\lab{dual-strength-rel} \\
\chi^2 = - \frac{2}{(p-1)^2}\, \g^{a_1 b_1}\ldots \g^{a_{p-1} b_{p-1}}
F_{a_1\ldots a_{p-1}}(\L) F_{b_1\ldots b_{p-1}}(\L) \; ,
\lab{chi-2}
\er
where $\chi \equiv \frac{\P (\vp)}{\sqrt{-\g}}$ is the variable brane tension, and:
\be
F_{a_1\ldots a_{p-1}}(\L) = (p-1) \pa_{[a_1} \L_{a_2\ldots a_{p-1}]}
\lab{dual-strength}
\ee
is the $(p-1)$-form dual field-strength.

All equations of motion can be equivalently derived from the following {\em dual} 
{\em WILL}-brane action:
\be
S_{\mathrm{dual}} \lb X,\g,\L\rb = - \h \int d^{p+1}\s\, \chi (\g,\L) \sqrt{-\g}
\g^{ab}\pa_a X^\m \pa_b X^\n G_{\m\n}(X)
\lab{WI-brane-dual}
\ee
with $\chi (\g,\L)$ given in \rf{chi-2} above.

\section{Intrinsically Light-Like Branes. {\em WILL}-Membrane}

Let us consider the $\g^{ab}$-equations of motion \rf{gamma-eqs}.
$F_{ab}$ is an anti-symmetric $(p+1)\times (p+1)$ matrix, therefore, $F_{ab}$ is 
{\em not invertible} in any odd $(p+1)$ -- it has at least one zero-eigenvalue 
vector $V^a$ ($F_{ab}V^b = 0$). Therefore, for any odd $(p+1)$ the induced
metric
\be
g_{ab} \equiv \(\pa_a X \pa_b X\) \equiv \pa_a X^\m \pa_b X^\n G_{\m\n}(X)
\lab{ind-metric}
\ee
on the world-volume of the Weyl-invariant brane \rf{WI-brane} is {\em singular} 
as {\em opposed} to the ordinary Nambu-Goto brane (where the induced metric
is proportional to the intrinsic Riemannian world-volume metric, cf. 
Eq.\rf{stand-brane-metric-eqs}):
\be
\(\pa_a X \pa_b X\) V^b = 0 \quad ,\quad \mathrm{i.e.}\;\;
\(\pa_V X \pa_V X\) = 0 \;\; ,\;\; \(\pa_{\perp} X \pa_V X\) = 0 \; ,
\lab{LL-constraints}
\ee
where $\pa_V \equiv V^a \pa_a$ and $\pa_{\perp}$ are derivates along the
tangent vectors in the complement of the tangent vector field $V^a$. 

Thus, we arrive at the following important conclusion:
every point on the world-surface of the Weyl-invariant $p$-brane \rf{WI-brane} 
(for odd $(p+1)$) moves with the speed of light in a time-evolution along the 
zero-eigenvalue vector-field $V^a$ of $F_{ab}$. Therefore, we will name 
\rf{WI-brane} (for odd $(p+1)$) by the acronym {\em WILL-brane} 
(Weyl-Invariant Light-Like-brane) model.

Henceforth we will explicitly consider the special case $p=2$ of \rf{WI-brane},
\textsl{i.e.}, the Weyl-invariant light-like membrane model.  
The associated {\em WILL}-membrane dual action (particular case of \rf{WI-brane-dual}
for $p=2$) reads:
\be
S_{\mathrm{dual}} = - \h \int d^3\s\, \chi (\g,u)\,\sqrt{-\g}
\g^{ab}\(\pa_a X \pa_b X\) \qquad ,\quad
\chi (\g,u) \equiv \sqrt{-2\g^{cd}\pa_c u \pa_d u} \; ,
\lab{WILL-membrane}
\ee
where $u$ is the dual ``gauge'' potential w.r.t. $A_a$:
\be
F_{ab}(A) = - \frac{1}{2\chi (\g,u)} \sqrt{-\g} \vareps_{abc} \g^{cd}\pa_d u\,
\g^{ef}\!\(\pa_e X \pa_f X\)  \; .
\lab{dual-strenght-rel-3}
\ee
$S_{\mathrm{dual}}$ is manifestly Weyl-invariant (under $\g_{ab} \to \rho\g_{ab}$).

The equations of motion w.r.t. $\g^{ab}$, $u$ (or $A_a$), and $X^\m$ read
accordingly:
\be
\(\pa_a X \pa_b X\) + \h \g^{cd}\(\pa_c X \pa_d X\) 
\(\frac{\pa_a u \pa_b u }{\g^{ef} \pa_e u \pa_f u} - \g_{ab}\) = 0 \; ,
\lab{gamma-eqs-3}
\ee
\be
\pa_a \(\,\frac{\sqrt{-\g}\g^{ab}\pa_b u}{\chi (\g,u)}\,
\g^{cd}\(\pa_c X \pa_d X\)\,\) = 0 \; ,
\lab{u-eqs}
\ee
\be
\pa_a \(\chi (\g,u)\,\sqrt{-\g} \g^{ab}\pa_b X^\m \) +
\chi (\g,u)\,\sqrt{-\g} \g^{ab}\pa_a X^\n \pa_b X^\l \G^\m_{\n\l} = 0 \; .
\lab{X-eqs-3}
\ee
The last factor in brackets on the l.h.s. of Eq.\rf{gamma-eqs-3} is a projector
implying that the induced metric $g_{ab} \equiv \(\pa_a X \pa_b X\)$ has 
zero-mode eigenvector $V^a =\g^{ab}\pa_b u$.

The invariance under world-volume reparametrizations allows to introduce the
following standard (synchronous) gauge-fixing conditions:
\be
\g^{0i} = 0 \;\; (i=1,2) \quad ,\quad \g^{00} = -1 \; .
\lab{gauge-fix}
\ee
Using \rf{gauge-fix} we can easily find solutions of Eq.\rf{u-eqs} for the
dual ``gauge potential'' $u$ in spite of its high non-linearity by taking the 
following ansatz:
\be
u (\t,\s^1,\s^2) = \frac{T_0}{\sqrt{2}}\t  \; ,
\lab{u-ansatz}
\ee
Here $T_0$ is an arbitrary integration constant with the dimension of membrane
tension. In particular:
\be
\chi \equiv \sqrt{-2\g^{ab}\pa_a u \pa_b u} = T_0 
\lab{chi-0}
\ee
The ansatz \rf{u-ansatz} means that we take $\t\equiv\s^0$ to be evolution
parameter along the zero-eigenvalue vector-field of the induced metric on the brane 
($V^a = \g^{ab}\pa_b u = \mathrm{const}\,(1,0,0)$).

  
The ansatz for $u$ \rf{u-ansatz} together with the gauge choice for $\g_{ab}$
\rf{gauge-fix} brings the equations of motion w.r.t. $\g^{ab}$, $u$ (or $A_a$) and 
$X^\m$ in the following form 
(recall $\(\pa_a X \pa_b X\) \equiv \pa_a X^\m \pa_b X^\n G_{\m\n}$):
\be
\(\pa_0 X \pa_0 X\) = 0 \quad ,\quad \(\pa_0 X \pa_i X\) = 0  \; ,
\lab{constr-0}
\ee
\be
\(\pa_i X\pa_j X\) - \h \g_{ij} \g^{kl}\(\pa_k X\pa_l X\) = 0  \; ,
\lab{constr-vir}
\ee
(notice that Eqs.\rf{constr-vir} look exactly like the classical (Virasoro) 
constraints for an Euclidean string theory with world-sheet parameters $(\s^1,\s^2)$);
\be
\pa_0 \(\sqrt{\g_{(2)}} \g^{kl}\(\pa_k X\pa_l X\)\) = 0  \; ,
\lab{u-eqs-fix}
\ee
where $\g_{(2)} = \det\Vert \g_{ij}\Vert$ 
(the above equation is the only remnant from the $A_a$-equations of motion \rf{A-eqs});
\be
\Box^{(3)} X^\m + \( - \pa_0 X^\n \pa_0 X^\l + 
\g^{kl} \pa_k X^\n \pa_l X^\l \) \G^{\m}_{\n\l} = 0  \; ,
\lab{X-eqs-3-fix}
\ee
where:
\be
\Box^{(3)} \equiv 
- \frac{1}{\sqrt{\g^{(2)}}} \pa_0 \(\sqrt{\g^{(2)}} \pa_0 \) + 
\frac{1}{\sqrt{\g^{(2)}}}\pa_i \(\sqrt{\g^{(2)}} \g^{ij} \pa_j \)   \; .
\lab{box-3}
\ee

We can also extend the {\em WILL}-brane model \rf{WI-brane} via a coupling 
to external space-time electromagnetic field $\cA_\m$. The natural
Weyl-conformal invariant candidate action reads (for $p=2$):
\be
S = - \int d^3\s \,\P (\vp) 
\Bigl\lb \h \g^{ab} \pa_a X^{\m} \pa_b X^{\n} G_{\m\n}
- \sqrt{F_{ab} F_{cd} \g^{ac}\g^{bd}}\Bigr\rb
- q\int d^3\s \, \vareps^{abc} \cA_\m \pa_a X^\m F_{bc} \; .
\lab{WILL-membrane+A}
\ee
The last Chern-Simmons-like term is a special case of a class of
Chern-Simmons-like couplings of extended objects to external electromagnetic
fields proposed in ref.\ct{Aaron-Eduardo}.

Instead of the action \rf{WILL-membrane+A} we can use its dual one (similar to 
the simpler case Eq.\rf{WI-brane} versus Eq.\rf{WILL-membrane}):
\be
S^{\mathrm{dual}}_{\mathrm{WILL-brane}}
= - \h \int d^3\s\, \chi (\g,u,\cA)\,\sqrt{-\g} \g^{ab}\(\pa_a X \pa_b X\)  \; ,
\lab{WILL-membrane+A-dual}
\ee
where the variable brane tension $\chi \equiv \frac{\P (\vp)}{\sqrt{-\g}}$ 
is given by:
\be
\chi (\g,u,\cA) \equiv \sqrt{-2\g^{cd}\(\pa_c u - q \cA_c\)
\(\pa_d u - q \cA_d\)} \quad ,\;\; \cA_a \equiv \cA_\m \pa_a X^\m \; .
\lab{tension+A}
\ee
Here $u$ is the dual ``gauge'' potential w.r.t. $A_a$ and the corresponding
field-strength and dual field-strength are related as
(cf. Eq.\rf{dual-strenght-rel-3}) :
\be
F_{ab}(A) = - \frac{1}{2\chi (\g,u,\cA)} \sqrt{-\g} \vareps_{abc} \g^{cd}
\(\pa_d u - q\cA_d \)\,\g^{ef}\!\(\pa_e X \pa_f X\)  \; .
\lab{dual-strenght-rel-3-A}
\ee

The corresponding equations of motion w.r.t. $\g^{ab}$, $u$ (or $A_a$), and $X^\m$
read accordingly:
\be
\(\pa_a X \pa_b X\) + \h \g^{cd}\(\pa_c X \pa_d X\)
\(\frac{\(\pa_a u - q\cA_a\)\(\pa_b u -q\cA_b\) }
{\g^{ef} \(\pa_e u - q\cA_e\)\(\pa_f u - q\cA_f\)} - \g_{ab}\) = 0  \; ;
\lab{gamma-eqs+A}
\ee
\be
\pa_a \(\,\frac{\sqrt{-\g}\g^{ab}\(\pa_b u - q\cA_b\)}{\chi (\g,u,\cA)}\,
\g^{cd}\(\pa_c X \pa_d X\)\,\) = 0  \; ;
\lab{u-eqs+A}
\ee
\br
\pa_a \(\chi (\g,u,\cA)\,\sqrt{-\g} \g^{ab}\pa_b X^\m \) +
\chi (\g,u,\cA)\,\sqrt{-\g} \g^{ab}\pa_a X^\n \pa_b X^\l \G^\m_{\n\l} 
\nonu \\
- \; q \vareps^{abc} F_{bc} \pa_a X^\n 
\(\pa_\l \cA_\n - \pa_\n \cA_\l\)\, G^{\l\m} = 0  \; . \phantom{aaaaaaaa}
\lab{X-eqs+A}
\er

\section{{\em WILL}-Membrane Solutions in Various Gravitational Backgrounds}    

\subsection{{\em WILL}-Membrane in a PP-Wave Background}
As a first non-trivial example let us consider \textsl{WILL}-membrane
dynamics in an external background generalizing the plane-polarized gravitational
wave (\textsl{pp-wave}):
\be
(ds)^2 = - dx^{+} dx^{-} - F(x^{+},x^I)\, (dx^{+})^2 + h_{IJ}(x^K) dx^I dx^J \; ,
\lab{gen-pp-wave}
\ee
(for the ordinary pp-wave $h_{IJ}(x^K) = \d_{IJ}$),
and let us employ in \rf{constr-0}--\rf{box-3} the following natural ansatz 
for $X^\m$ (here $\s^0 \equiv \t$; $I=1,\ldots,D-2$) :
\be
X^{-} = \t \quad, \quad 
X^{+}=X^{+}(\t,\s^1,\s^2) \quad, \quad X^I = X^I (\s^1,\s^2) \; .
\lab{ansatz-pp-wave}
\ee
The non-zero affine connection symbols for the generalized pp-wave metric 
\rf{gen-pp-wave} are: $\G^{-}_{++}=\pa_{+}F$, $\G^{-}_{+I}=\pa_{I}F$, 
$\G^{I}_{++}=\h h^{IJ}\pa_{J}F$, and $\G^{I}_{JK}$ -- the ordinary Cristoffel
symbols for the metric $h_{IJ}$ in the transverse dimensions.

It is straightforward to show that the solution does not depend on the form of 
the pp-wave front $F(x^{+},x^I)$ and reads:
\be
X^{+}=X^{+}_0 = \mathrm{const} \quad ,\quad 
\g_{ij} \; \mathrm{are}\; \t\!-\!\mathrm{independent}\; ;
\lab{pp-wave-sol-1}
\ee
\be
\(\pa_i X^I \pa_j X^J - \h \g_{ij} \g^{kl} \pa_k X^I \pa_l X^J\)\, h_{IJ} = 0 
\lab{pp-wave-sol-2}
\ee
\be
\frac{1}{\sqrt{\g^{(2)}}} \pa_i \(\sqrt{\g^{(2)}} \g^{ij} \pa_j X^I \) 
+ \g^{kl} \pa_k X^J \pa_l X^K \G^{I}_{JK} = 0 
\lab{pp-wave-sol-3}
\ee
The latter two equations for the transverse brane coordinates describe a string 
moving in the $(D-2)$-dimensional Euclidean-signature transverse space.

\subsection{{\em WILL}-Membrane in a Product-Space Background}

Here we consider {\em WILL}-membrane 
moving in a general product-space $D=(d+2)$-dimensional gravitational
background $\cM^d \times \S^{2}$ with coordinates $(x^\m, y^m)$ ($\m =
0,1,\ldots ,d-1$, $m = 1,2$) and Riemannian metric
$(ds)^2 = f(y) g_{\m\n}(x) dx^\m dx^\n + g_{mn}(y) dy^m dy^n$. The metric on
$\cM^d$ is of Lorentzian signature and $\S^{2}$ will be taken as a sphere
for simplicity.

We assume that the {\em WILL}-brane wraps around the ``internal''
space $\S^{2}$ and use the following ansatz (recall $\t \equiv \s^0$):
\be
X^\m = X^\m (\t) \quad , \quad Y^m = \s^m \quad , \quad
\g_{mn} = a(\t)\, g_{mn}(\s^1,\s^2)
\lab{ansatz-product-space}
\ee
Then the equations of motion and constraints \rf{constr-0}--\rf{box-3}
reduce to:
\be
\pa_\t X^\m \pa_\t X^\n g_{\m\n}(X) = 0 \quad ,\quad
\frac{1}{a(\t)}\,\pa_\t \Bigl( a(\t) \pa_\t X^\m \Bigr) + 
\pa_\t X^\n \pa_\t X^\l\, \G^\m_{\n\l} = 0 
\lab{eff-massless}
\ee
where $a(\t)$ is the conformal factor of the space-like part of the internal
membrane metric (last Eq.\rf{ansatz-product-space}). 
Eqs.\rf{eff-massless} are of the same form
as the equations of motion for a massless point-particle with a world-line
``einbein'' $e = a^{-1}$ moving in $\cM^d$. In other words, the simple solution 
above describes a membrane living in the extra ``internal'' dimensions $\S^{2}$ and
moving as a whole with the speed of light in ``ordinary'' space-time $\cM^d$.

Let us particularly emphasize the fact that, although the {\em WILL}-brane 
is wrapping the extra (compact) dimensions in a topologically non-trivial way 
(cf. second Eq.\rf{ansatz-product-space}), its modes remain {\em massless} from the
projected $d$-dimensional space-time point of view. This is a new phenomenon
from the point of view of Kaluza-Klein theories: here we have particles 
(membrane modes), which aquire non-zero quantum numbers due to non-trivial winding, 
while at the same time these particles (modes) remain massless. 
In contrast, one should recall that in ordinary Kaluza-Klein
theory (for a review, see \ct{K-K-review}), 
non-trivial dependence on the extra dimensions is possible for
point particles or even standard strings and branes only at a very high
energy cost (either by momentum modes or winding modes), which implies a
very high mass from the projected $d$-dimensional space-time point of view. 


\subsection{{\em WILL}-Membrane in Spherically-Symmetric Backgrounds}

Let us consider general $SO(3)$-symmetric background in $D=4$ embedding space-time:
\be
(ds)^2 =  - A(z,t) (dt)^2 + B(z,t) (dz)^2 + C (z,t) \( (d\th)^2 + \sin^2\th (d\p)^2\).
\lab{backgr}
\ee

The usual ansatz:
\br
X^0 \equiv t = \t \quad ,\quad
X^1 \equiv z = z (\t,\s^1,\s^2) \quad ,\quad
X^2 \equiv \th = \s^1 \quad ,\quad X^3 \equiv \p = \s^2
\lab{so3-ansatz} \\
\g_{ij} = a (\t) \( (d\s^1)^2 + \sin^2(\s^1) (d\s^2)^2\)
\phantom{aaaaaaaaaaaaa}
\nonu
\er
yields: 

(i) equations for $z (\t,\s^1,\s^2)$ :
\be
\frac{\pa z}{\pa \t} = \pm \sqrt{\frac{A}{B}}  \quad ,\quad 
\frac{\pa z}{\pa \s^i} = 0
\;\; ;
\lab{z-eqs}
\ee

(ii) a restriction on the background itself (comes from the gauge-fixed equations of 
motion for the dual gauge potential $u$ \rf{u-eqs-fix}) :
\be
\frac{d C}{d\t} \equiv \(\partder{C}{t} \pm 
\sqrt{\frac{A}{B}}\, \partder{C}{z}\)\bgv_{t=\t,\; z=z(\t)} = 0   \;\; ;
\lab{C-eq}
\ee

(iii) an equation for the conformal factor $a(\t)$ of the internal membrane metric:
\be
\pa_\t a + \(\frac{\partder{}{t}\sqrt{AB} \pm \pa_z A}{\sqrt{AB}}
\bgv_{t=\t,\, z=z(\t)}\)\, a(\t)
- \frac{\partder{}{t}C}{A}\bgv_{t=\t,\, z=z(\t)} = 0 \;\; .
\lab{a-eq}
\ee
Eq.\rf{C-eq} tells that the (squared) sphere radius $R^2 \equiv C (z,t)$ must remain
constant along the {\em WILL}-brane trajectory. 




In particular, let us take static spherically-symmetric gravitational
background in $D=4$:
\be
(ds)^2 = - A(r)(dt)^2 + B(r)(dr)^2 + r^2 \lb (d\th)^2 + \sin^2 (\th)\,(d\p)^2\rb \; .
\lab{spherical-symm-metric}
\ee
Specifically we have:
\be
A(r) = B^{-1}(r) = 1 - \frac{2GM}{r}
\lab{schwarzschild}
\ee
for Schwarzschild black hole,
\be
A(r) = B^{-1}(r) = 1 - \frac{2GM}{r} + \frac{Q^2}{r^2}
\lab{R-N}
\ee
for Reissner-Nordstr\"{o}m black hole,
\be
A(r) = B^{-1}(r) = 1 - \k r^2
\lab{AdS}
\ee
for (anti-) de Sitter space, \textsl{etc.}.

In the case of \rf{spherical-symm-metric} Eqs.\rf{z-eqs}--\rf{C-eq} reduce to:
\be
\partder{r}{\t} = \pm A(r) \quad ,\quad \partder{r}{\s^i} = 0 \quad ,\quad
\partder{r}{\t} = 0 
\lab{sol-spherical-1}
\ee
yielding:
\be
r = r_0 \equiv \mathrm{const} \;\;,\;\; \mathrm{where} \quad A(r_0)=0  \; .
\lab{sol-spherical-2}
\ee
Further, Eq.\rf{a-eq} implies for the intrinsic {\em WILL}-membrane metric:
\be
\Vert\g_{ij}\Vert = c_0 \, e^{\mp \t/r_0}\,\twomat{1}{0}{0}{\sin^2 (\s^1)} \; ,
\lab{sol-spherical-3}
\ee
where $c_0$ is an arbitrary integration constant.

From \rf{sol-spherical-2} we conclude that the {\em WILL}-membrane with spherical 
topology (and with exponentially blowing-up/deflating radius w.r.t. internal
metric, see Eq.\rf{sol-spherical-3}) automatically ``sits'' on (materializes) 
the event horizon of the pertinent black hole in $D=4$ embedding space-time. This 
conforms with the well-known general property of closed light-like hypersurfaces 
in $D=4$ (\textsl{i.e.}, their section with the hyper-plane $t$=const~ being a 
compact 2-dimensional manifold) which automatically serve as horizons \ct{LL-vol2}. 
On the other hand, let us stress that our \textsl{WILL}-membrane model 
\rf{WILL-membrane} provides an explicit {\em dynamical} realization of event horizons.

\section{Coupled Einstein-Maxwell-{\em WILL}-Membrane System: 
{\em WILL}-Membrane as a Source for Gravity and Electromagnetism}

We can extend the results from the previous section to the case of the 
self-consistent Einstein-Maxwell-{\em WILL}-membrane system, \textsl{i.e.}, we will
consider the {\em WILL}-membrane as a dynamical material and electrically 
charged source for gravity and electromagnetism. The relevant action reads:
\be
S = \int\!\! d^4 x\,\sqrt{-G}\,\llb \frac{R (G)}{16\pi G_N}
- \frac{1}{4} \cF_{\m\n}(\cA) \cF_{\k\l}(\cA) G^{\m\k} G^{\n\l}\rrb
+ S_{\mathrm{WILL-brane}}  \; ,
\lab{E-M-WILL} 
\ee
where $\cF_{\m\n}(\cA) = \pa_\m \cA_\n - \pa_\n \cA_\m$ is the space-time
electromagnetic field-strength, and $S_{\mathrm{WILL-brane}}$ indicates the
the \textsl{WILL}-membrane action coupled to the space-time gauge field $\cA_\m$
-- either \rf{WILL-membrane+A} or its dual \rf{WILL-membrane+A-dual}.

The equations of motion for the \textsl{WILL}-membrane subsystem are of the same 
form as Eqs.\rf{gamma-eqs+A}--\rf{X-eqs+A}. The Einstein-Maxwell equations of motion
read:
\be
R_{\m\n} - \h G_{\m\n} R = 8\pi G_N \( T^{(EM)}_{\m\n} + T^{(brane)}_{\m\n}\)\; ,
\lab{Einstein-eqs}
\ee
\be
\pa_\n \(\sqrt{-G}G^{\m\k}G^{\n\l} \cF_{\k\l}\) + j^\m = 0 \; ,
\lab{Maxwell-eqs}
\ee
where:
\be
T^{(EM)}_{\m\n} \equiv \cF_{\m\k}\cF_{\n\l} G^{\k\l} - G_{\m\n}\frac{1}{4}
\cF_{\r\k}\cF_{\s\l} G^{\r\s}G^{\k\l} \; ,
\lab{T-EM}
\ee
\be
T^{(brane)}_{\m\n} \equiv - G_{\m\k}G_{\n\l}
\int\!\! d^3 \s\, \frac{\d^{(4)}\Bigl(x-X(\s)\Bigr)}{\sqrt{-G}}\,
\chi\,\sqrt{-\g} \g^{ab}\pa_a X^\k \pa_b X^\l \; ,
\lab{T-brane}
\ee
\be
j^\m \equiv q \int\!\! d^3 \s\,\d^{(4)}\Bigl(x-X(\s)\Bigr)
\vareps^{abc} F_{bc} \pa_a X^\m \; .
\lab{brane-EM-current}
\ee

We find the following self-consistent spherically symmetric stationary solution 
for the coupled Einstein-Maxwell-{\em WILL}-membrane system \rf{E-M-WILL}. 
For the Einstein subsystem we have a solution:
\be
(ds)^2 = - A(r)(dt)^2 + A^{-1}(r)\,(dr)^2 + 
r^2 \lb (d\th)^2 + \sin^2 (\th)\,(d\p)^2\rb  \; ,
\lab{spherical-symm-metric-b}
\ee
consisting of two different black holes with a {\em common} event horizon:
\begin{itemize}
\item    
Schwarzschild black hole inside the horizon:
\be
A(r)\equiv A_{-}(r) = 1 - \frac{2GM_1}{r}\;\; ,\quad \mathrm{for}\;\; 
r < r_0 \equiv r_{\mathrm{horizon}}= 2GM_1 \; .
\lab{Schwarzschild-metric-in}
\ee
\item
Reissner-Norstr\"{o}m black hole outside the horizon:
\be
A(r)\equiv A_{+}(r) = 1 - \frac{2GM_2}{r} + \frac{GQ^2}{r^2}\;\; ,
\quad \mathrm{for}\;\; r > r_0 \equiv r_{\mathrm{horizon}} \; ,
\lab{RN-metric-out}
\ee
where $Q^2 = 8\pi q^2 r_{\mathrm{horizon}}^4 \equiv 128\pi q^2 G^4 M_1^4$;
\end{itemize}
For the Maxwell subsystem we have $\cA_1 = \ldots =\cA_{D-1}=0$ everywhere and:
\begin{itemize}
\item
Coulomb field outside horizon:
\be
\cA_0 = \frac{\sqrt{2}\, q\, r_{\mathrm{horizon}}^2}{r} \;\; ,\quad \mathrm{for}\;\; 
r \geq r_0 \equiv r_{\mathrm{horizon}}  \; .
\lab{EM-out}
\ee
\item
No electric field inside horizon:
\be
\cA_0 = \sqrt{2}\, q\, r_{\mathrm{horizon}} = \mathrm{const} \;\; ,\quad \mathrm{for}\;\; 
r \leq r_0 \equiv r_{\mathrm{horizon}}  \; .
\lab{EM-in}
\ee
\end{itemize}

Using the same (synchronous) gauge choice \rf{gauge-fix} and ansatz for the dual
``gauge potential'' \rf{u-ansatz}, as well as taking into account 
\rf{EM-out}--\rf{EM-in}, the \textsl{WILL}-membrane equations of motion
\rf{gamma-eqs+A}--\rf{X-eqs+A} acquire the form
(recall $\(\pa_a X \pa_b X\) \equiv \pa_a X^\m \pa_b X^\n G_{\m\n}$):
\be
\(\pa_0 X \pa_0 X\) = 0 \quad ,\quad \(\pa_0 X \pa_i X\) = 0  \; ,
\lab{constr-0+A}
\ee
\be
\(\pa_i X\pa_j X\) - \h \g_{ij} \g^{kl}\(\pa_k X\pa_l X\) = 0 \; ,
\lab{constr-vir+A}
\ee
(these constraints are the same as in the absence of coupling to space-time
gauge field \rf{constr-0}--\rf{constr-vir});
\be
\pa_0 \(\sqrt{\g_{(2)}} \g^{kl}\(\pa_k X\pa_l X\)\) = 0 \; ,
\lab{u-eqs-fix+A}
\ee
(once again the same equation as in the absence of coupling to space-time
gauge field \rf{u-eqs-fix});
\be
{\wti \Box}^{(3)} X^\m + \( - \pa_0 X^\n \pa_0 X^\l + 
\g^{kl} \pa_k X^\n \pa_l X^\l \) \G^{\m}_{\n\l} 
- q \frac{\g^{kl}\(\pa_k X \pa_l X\)}{\sqrt{2}\,\chi}\,
\pa_0 X^\n \(\pa_\l \cA_\n - \pa_\n \cA_\l \) G^{\l\m} = 0 \; .
\lab{X-eqs-fix+A}
\ee
Here $\chi \equiv T_0 - \sqrt{2}q\cA_0$ 
with $\cA_0$ as in Eqs.\rf{EM-out},\rf{EM-in}
is the variable brane tension coming from Eqs.\rf{u-ansatz},\rf{tension+A};
$X^0 \equiv t, X^1 \equiv r, X^2 \equiv \th, X^3 \equiv \phi$; and:
\be
{\wti \Box}^{(3)} \equiv 
- \frac{1}{\chi \sqrt{\g^{(2)}}} \pa_0 \(\chi \sqrt{\g^{(2)}} \pa_0 \) + 
\frac{1}{\chi \sqrt{\g^{(2)}}}\pa_i \(\chi \sqrt{\g^{(2)}} \g^{ij} \pa_j \)
\; .
\lab{box-3+A}
\ee

A self-consistent solution to Eqs.\rf{constr-0+A}--\rf{X-eqs-fix+A} reads:
\be
X^0 \equiv t = \t \quad,\quad \th = \s^1 \quad,\quad \p = \s^2 \;\; , 
\lab{Schwarzschild-RN-sol-1}
\ee
\be
r (\t,\s^1,\s^2) = r_{\mathrm{horizon}} = \mathrm{const} \quad,\quad 
A_{\pm}(r_{\mathrm{horizon}})=0 \; ,
\lab{Schwarzschild-RN-sol-2}
\ee
\textsl{i.e.}, the \textsl{WILL}-membrane automatically positions itself on the 
common event horizon of the pertinent black holes. Furthermore, inserting 
\rf{Schwarzschild-RN-sol-1}--\rf{Schwarzschild-RN-sol-2} in the expression 
\rf{T-brane} for the \textsl{WILL}-membrane energy-momentum tensor 
$T^{(brane)}_{\m\n}$, the Einstein equations \rf{Einstein-eqs} entail the 
following important matching conditions for the space-time metric components 
along the \textsl{WILL}-membrane surface:
\be
\partder{}{r} A_{+}\bgv_{r=r_{\mathrm{horizon}}} -
\partder{}{r} A_{-}\bgv_{r=r_{\mathrm{horizon}}} = - 16\pi G \chi \; .
\lab{metric-match}
\ee
Condition \rf{metric-match} in turn yields relations between the parameters of the 
black holes and the \textsl{WILL}-membrane ($q$ being its surface charge density) :
\be
M_2 = M_1 + 32\pi q^2 G^3 M_1^3
\lab{mass-match}
\ee
and for the brane tension $\chi$:
\be
\chi \equiv T_0 - 2q^2r_{\mathrm{horizon}} =q^2 G M_1 \quad, \;\;
\mathrm{i.e.} \;\; T_0 = 5 q^2 G M_1
\lab{tension-match}
\ee

The matching condition \rf{metric-match} corresponds to the so called statically
soldering conditions in the theory of light-like thin shell dynamics in general 
relativity \ct{Barrabes-Israel}. Unlike the latter model, where the membranes are 
introduced {\em ad hoc}, the present {\em WILL}-brane models provide a systematic
dynamical description of light-like branes (as sources for both gravity and
electromagnetism) from first principles starting with concise Weyl-conformally
invariant actions \rf{WILL-membrane+A}, \rf{E-M-WILL}. 

\section{Conclusions and Outlook} 

In the present work we have discussed a novel class of Weyl-invariant $p$-brane 
theories whose dynamics significantly differs from ordinary Nambu-Goto $p$-brane
dynamics. The princial features of our construction are as follows:
\begin{itemize}
\item
Employing alternative non-Riemannian integration measure (volume-form) 
\rf{mod-measure-p} on the $p$-brane world-volume independent of the intrinsic 
Riemannian metric.
\item
Acceptable dynamics in the novel class of brane models
(Eqs.\rf{WI-brane},\rf{WILL-membrane+A}) {\em naturally} requires the 
introduction of additional world-volume gauge fields.
\item
By employing square-root Yang-Mills actions for the pertinent
world-volume gauge fields one achieves manifest {\em Weyl-conformal
symmetry} in the new class of $p$-brane theories {\em for any $p$}.
\item
The brane tension is {\em not} a constant dimensionful scale given 
{\em ad hoc}, but rather it appears as a {\em composite} world-volume scalar 
field (Eqs.\rf{chi-2},\rf{WILL-membrane},\rf{tension+A}) 
transforming non-trivially under Weyl-conformal transformations.
\item
The novel class of Weyl-invariant $p$-brane theories describes
intrinsically {\em light-like} $p$-branes for any even $p$ ({\em WILL}-branes).
\item
When put in a gravitational black hole background, the {\em WILL}-membrane
($p=2$) automatically sits on (``materializes'') the event horizon.
\item
When moving in background product-spaces (``Kaluza-Klein'' context) the 
\textsl{WILL}-membrane describes {\em massless} modes, even though the
membrane is wrapping the extra dimensions and therefore aquiring non-trivial 
Kaluza-Klein charges.
\item
The coupled Einstein-Maxwell-{\em WILL}-membrane system \rf{E-M-WILL} possesses
self-consistent solution where the {\em WILL}-membrane serves as a 
material and electrically charged source for gravity and electromagnetism,
and it automatically ``sits'' on (materializes) the common
event horizon for a Schwarzschild (in the interior) and Reissner-Nordstr\"{o}m
(in the exterior) black holes. Thus our model \rf{E-M-WILL} provides an
explicit dynamical realization of the so called ``membrane paradigm'' in the
physics of black holes \ct{membrane-paradigm}. 
\item
The {\em WILL}-branes could be good representations for the string-like objects
introduced by 't Hooft in ref.\ct{Hooft} to describe gravitational interactions
associated with black hole formation and evaporation, since as shown above the 
{\em WILL}-branes locate themselves automatically in the horizons and,
therefore, they could represent degrees of freedom associated particularly with 
horizons.
\end{itemize}

The novel class of Weyl-conformal invariant $p$-branes discussed above
suggests various physically interesting directions for further study such as:
quantization (Weyl-conformal anomaly and critical dimensions); supersymmetric
generalization; possible relevance for the open string dynamics (similar to the
role played by Dirichlet- ($Dp$-)branes); {\em WILL}-brane dynamics in more 
complicated gravitational black hole backgrounds (\textsl{e.g.}, Kerr-Newman).

\vspace{.1in}

\textbf{Acknowledgements.}
Two of us (E.N. and S.P.) are sincerely grateful for hospitality and
support to Prof. Branko Dragovich and the organizers of
the \textsl{Third Summer School on Modern Mathematical Physics} (Zlatibor, Serbia
and Montenegro, August 2004), and to the organizers of the \textsl{Second 
Annual Meeting of the European RTN}~ {\em EUCLID} (Sozopol, Bulgaria, September 2004).
One of us (E.G.) thanks the Institute for Nuclear Research and Nuclear Energy
(Sofia) and Trieste University for hospitality. He also acknowledges useful 
conversations with Gerard `t Hooft, Euro Spallucci and Stefano Ansoldi.

E.N. and S.P. are partially supported by Bulgarian NSF grant \textsl{F-1412/04}
and European RTN {\em ``Forces-Universe''} (Contract No.\textsl{MRTN-CT-2004-005104}).
Finally, all of us acknowledge support of our collaboration through the exchange
agreement between the Ben-Gurion Univesity of the Negev (Beer-Sheva, Israel) and
the Bulgarian Academy of Sciences.

    
\end{document}